\documentclass{moriond}

\def\be{\begin{equation}}
\def\ee{\end{equation}}
\def\bea{\begin{eqnarray}}
\def\eea{\end{eqnarray}}

\usepackage{xspace}

\newcommand{\lhcb}{LHCb\xspace}
\newcommand{\pythia}{\textsc{Pythia}\xspace}
\newcommand{\evtgen}{\textsc{EvtGen}\xspace}
\newcommand{\photos}{\textsc{Photos}\xspace}
\newcommand{\geant}{\textsc{Geant4}\xspace}
\newcommand{\sPlot}{\mbox{\em sPlot}\xspace}

\newcommand{\CP}{\ensuremath{C\!P}\xspace}

\newcommand{\Lz}{\ensuremath{\Lambda}\xspace}

\newcommand{\Xibz}{\ensuremath{\Xi_b^0}\xspace}
\newcommand{\Lb}{\ensuremath{\Lambda_b^0}\xspace}

\newcommand{\invfb}{\ensuremath{\mathrm{fb}^{-1}}\xspace}
\newcommand{\gev}{\ensuremath{\mathrm{GeV}}\xspace}
\newcommand{\mev}{\ensuremath{\mathrm{MeV}}\xspace}

\newcommand{\Photo}{}

\begin{document}
\vspace*{4cm}

\title{OBSERVATION OF THE CHARMLESS PURELY BARYONIC DECAY $\Lb \to \Lz p \bar{p}$}

\author{A. Brea Rodriguez, on behalf of the LHCb collaboration}

\address{École polytechnique fédérale de Lausanne (EPFL), LPHE, Lausanne, Switzerland}

\maketitle\abstracts{
The first observation of a charmless purely baryonic decay,
$\Lb \to \Lz p \bar{p}$, is reported using the full Run~2 \lhcb dataset, corresponding to an integrated luminosity of $6.0~\invfb$.
The branching fraction is measured relative to that of the topologically similar
normalisation mode $\Lb \to \Lz K^+K^-$.
A simultaneous fit to the long- and downstream-track categories yields a signal
significance of $5.1\sigma$ after including systematic uncertainties.
The relative branching fraction is measured to be
$\left(5.13 \pm 1.28_{\rm stat} \pm 0.27_{\rm syst}\right)\times 10^{-2}$
in the region $m(h^+h^-)<2.85~\gev$.
}

\section{Introduction}

Purely baryonic decays of beauty hadrons provide a novel laboratory to study
the dynamics of multibody baryonic final states and to search for
\CP-violating effects in systems with rich spin structure.
While several baryonic $B$-meson decays have been observed in recent
years~\cite{PDG2024,LHCb-PAPER-2017-005,LHCb-PAPER-2017-022,LHCb-PAPER-2022-032,LHCb-PAPER-2025-032,LHCb-PAPER-2025-053},
purely baryonic decays remain largely unexplored.
So far, the only experimentally established purely baryonic modes are
$\Lb \to \Sigma_c^+ \bar{p} p$ and $\Lb \to \Sigma_c^{*+}\bar{p} p$,
observed by the \lhcb collaboration in the study of
$\Lb \to \Lambda_c^+ \bar{p}\pi^- $ decays~\cite{LHCb-PAPER-2018-005}.

This contribution presents the first dedicated search for a purely baryonic decay of a baryon and the first observation of a charmless such decay, $\Lb \to \Lz p \bar{p}$, using the full Run~2 \lhcb dataset.
The branching fraction is measured relative to that of the topologically
similar decay $\Lb \to \Lz K^+ K^-$.
Theoretical calculations predict
\begin{equation}
\mathcal{B}(\Lb \to \Lz p \bar{p}) =
(3.2^{+0.8}_{-0.3}\pm0.4\pm0.7)\times10^{-6},
\end{equation}
where the quoted uncertainties are associated with nonfactorisable effects,
CKM matrix elements, and hadronic form factors, respectively~\cite{PBD1}.
The same framework also predicts sizeable direct \CP asymmetries in
$\Lb \to \Lz p \bar{p}$ and $\Xibz \to \Lz p \bar{p}$ decays~\cite{PBD1}.

\section{Detector, dataset and analysis strategy}

The \lhcb detector is a single-arm forward spectrometer covering the
pseudorapidity range $2<\eta<5$, designed for the study of hadrons containing
$b$ or $c$ quarks~\cite{LHCb-DP-2008-001,LHCb-DP-2014-002}. Its main
subsystems relevant for this analysis are a high-precision tracking system,
two ring-imaging Cherenkov detectors for hadron identification, calorimeters,
and a muon system. The online event selection is performed by a hardware
trigger followed by a software trigger~\cite{LHCb-DP-2012-004}.

The analysis uses the full Run~2 data sample collected in 2015--2018,
corresponding to an integrated luminosity of $6.0\invfb$.
The \Lz baryon is reconstructed through the decay $\Lz \to p \pi^-$ in two
categories: \emph{long} (LL), when both decay-product tracks are reconstructed in
the vertex detector, and \emph{downstream} (DD), when the \Lz decays outside
its acceptance. About half of the selected signal candidates belong to the DD
category, so both are retained.

Simulated samples are used to determine efficiencies, validate the fit model,
and study background contributions. In the simulation, $pp$ collisions are
generated using \pythia~\cite{Sjostrand:2007gs} with a
specific \lhcb configuration~\cite{LHCb-PROC-2010-056}. Hadron decays are
described by \evtgen~\cite{Lange:2001uf}, with final-state radiation generated
using \photos~\cite{davidson2015photos}. The interaction with the detector is
implemented using \geant~\cite{Allison:2006ve}.
The three-body signal and normalisation decays are generated uniformly in the
square-Dalitz variables~\cite{PhysRevD.72.052002}.

The main observable is the ratio
\begin{equation}
R_{\Lb}\equiv
\frac{\mathcal{B}(\Lb \to \Lz p \bar{p})}
     {\mathcal{B}(\Lb \to \Lz K^+ K^-)}
=
\frac{N(\Lb \to \Lz p \bar{p})}
     {N(\Lb \to \Lz K^+ K^-)}
\cdot
\frac{\epsilon_{\Lb \to \Lz K^+ K^-}}
     {\epsilon_{\Lb \to \Lz p \bar{p}}},
\label{eq:ratio}
\end{equation}
where $N$ denotes the fitted yields and $\epsilon$ the corresponding total
efficiencies, including detector acceptance, trigger, reconstruction and
selection effects. The use of the topologically similar
$\Lb \to \Lz K^+ K^-$ decay as normalisation strongly reduces systematic
uncertainties in the ratio.

To avoid experimenter's bias, the results of the analysis weren't examined until the full procedure had been finalised. Candidates with reconstructed
$\Lambda p\bar p$ invariant mass within $\pm50$  \mev of the known \Lb and
\Xibz masses~\cite{PDG2024} are excluded until the selection, fit model, and systematic
studies are finalised. In addition to the $\Lb \to \Lz p \bar{p}$ signal, the
fit includes a possible contribution from $\Xibz \to \Lz p \bar{p}$ decays,
which would appear about 175 $\mev$ above the \Lb peak.

\section{Selection, efficiencies and fit model}

The event selection follows a common strategy for the signal and normalisation
modes. It combines trigger, offline reconstruction, preselection, a dedicated
\textsc{XGBoost} classifier~\cite{XGBoosting}, particle-identification (PID) requirements on the companion hadrons, and
charm-hadron vetoes. In both channels, a requirement
$m(h^+h^-)<2.85$ \gev is imposed to exclude the charmonium region. In the
normalisation mode, additional vetoes suppress intermediate charm-hadron
contributions and kaon-pion misidentification backgrounds. The multivariate classifier is trained using simulated
$\Lb \to \Lz K^+ K^-$ decays as signal proxy and candidates from the upper
sideband of the $m(\Lz p \bar{p})$ distribution in data as background.
Separate classifiers are trained for the LL and DD categories.
The same classifier response is used for the signal and normalisation modes,
while the PID requirement is optimised separately for proton and kaon
pairs.

The relative efficiencies are corrected for known data-simulation
differences in tracking, trigger, and PID response using standard \lhcb
calibration procedures~\cite{LHCb-DP-2013-002,LHCb-DP-2018-001}. Since the
kinematics of the two decay modes are very similar, most effects cancel in the
efficiency ratio. To account for nonuniform populations of the three-body
phase space, efficiency maps are constructed from simulation and then folded
with the background-subtracted Dalitz distributions observed in data.
The latter are obtained with the \sPlot technique~\cite{Pivk:2004ty}, using
the invariant mass as discriminating variable.
This procedure is applied to both the signal and normalisation modes and yields
phase-space-averaged efficiency ratios separately for the LL and DD
categories. The efficiency ratios differ by about 14\% from those obtained
under the assumption of uniform phase space, showing that this correction is
important for the measurement.

\begin{figure}[t]
\begin{minipage}{0.44\linewidth}
\centerline{\includegraphics[width=\linewidth]{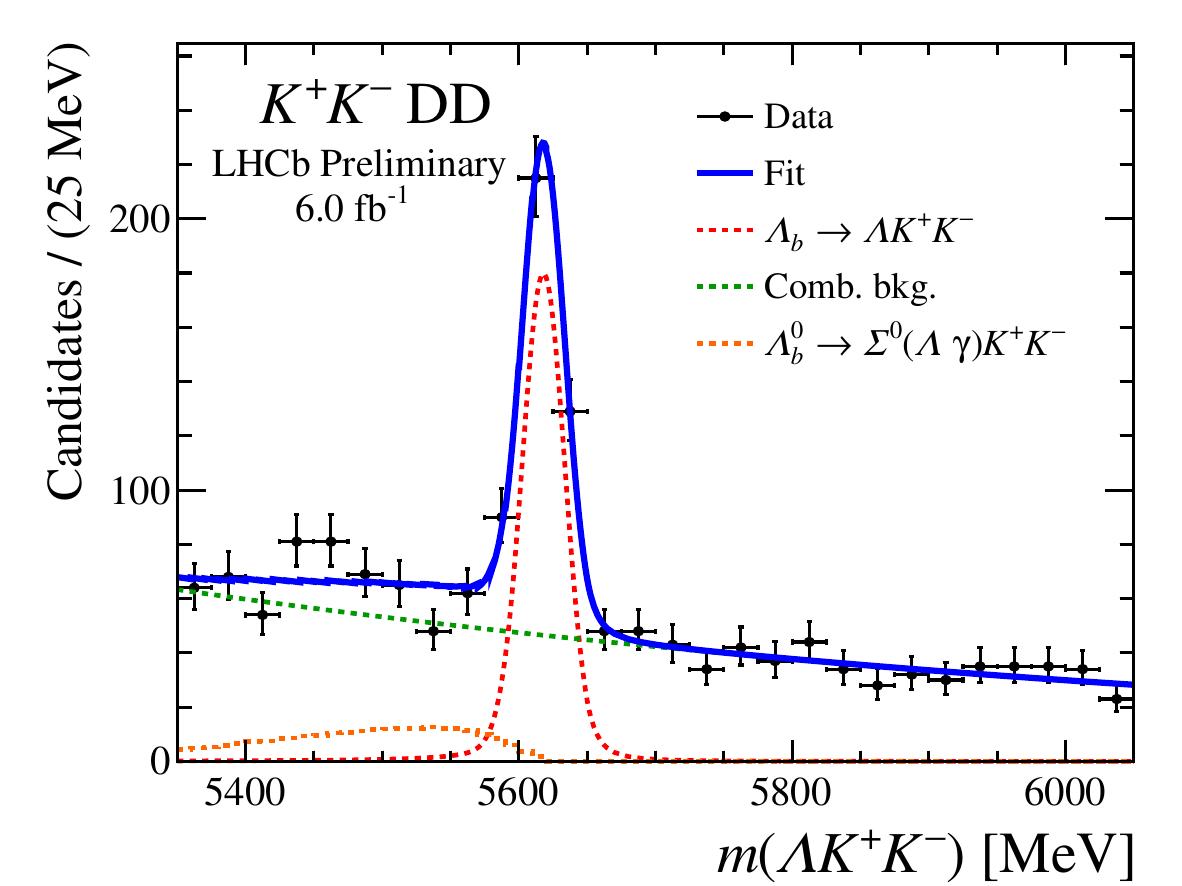}}
\end{minipage}
\hfill
\begin{minipage}{0.44\linewidth}
\centerline{\includegraphics[width=\linewidth]{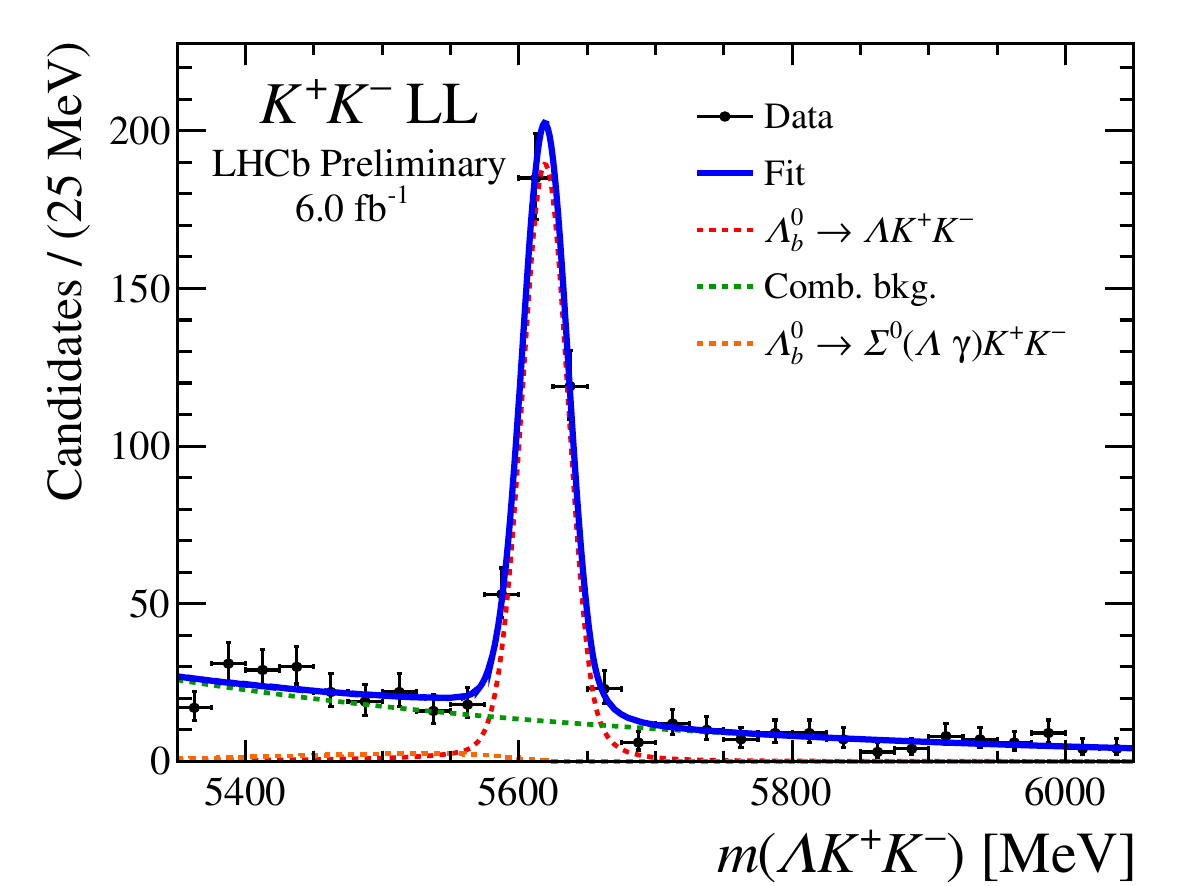}}
\end{minipage}

\vspace{0.2cm}

\begin{minipage}{0.44\linewidth}
\centerline{\includegraphics[width=\linewidth]{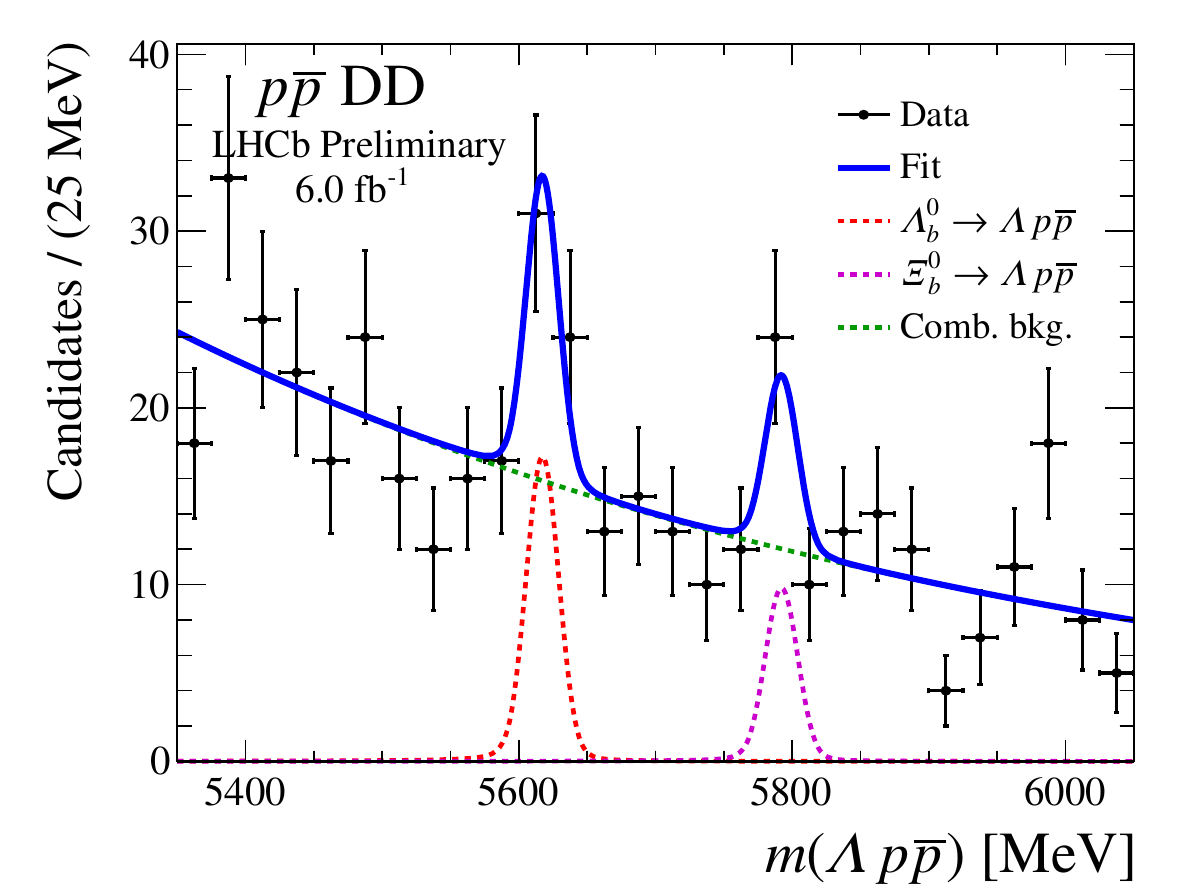}}
\end{minipage}
\hfill
\begin{minipage}{0.44\linewidth}
\centerline{\includegraphics[width=\linewidth]{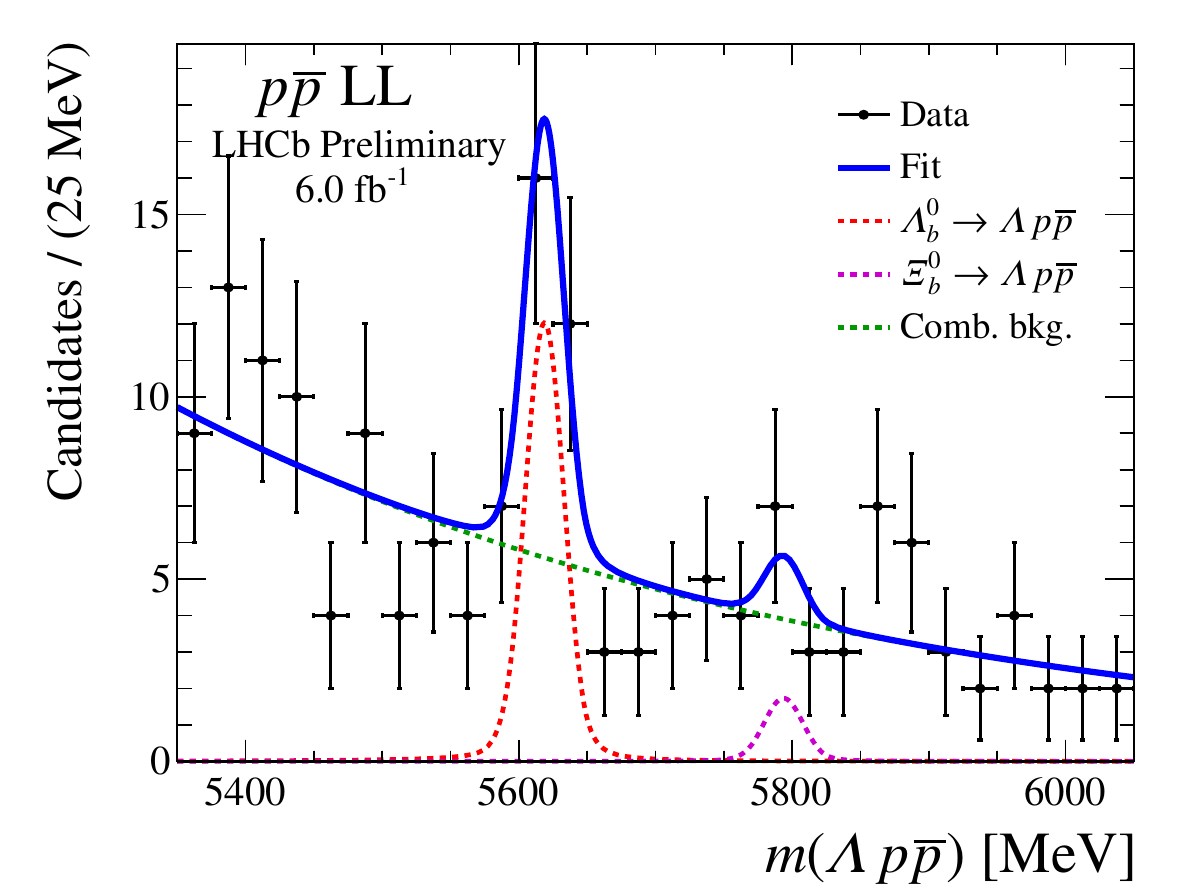}}
\end{minipage}
\caption[]{Invariant-mass distributions of (top) $\Lb \to \Lz K^+K^-$ and (bottom) $\Lb \to \Lz p\bar{p}$ candidates in the (left) DD and (right) LL categories in Run~2 data with the full selection applied. The fit model is shown as a solid line.}
\label{fig:SignalFits}
\end{figure}

The signal yields are obtained from an extended unbinned
maximum-likelihood fit (see Fig.~\ref{fig:SignalFits}) performed simultaneously to the four invariant-mass
spectra corresponding to the $\Lb \to \Lz K^+K^-$ and
$\Lb \to \Lz p \bar{p}$ modes in the LL and DD categories.
In the normalisation mode, the fit includes the $\Lb \to \Lz K^+K^-$ signal,
a partially reconstructed $\Lb \to \Sigma^0 K^+K^-$ contribution, and
combinatorial background.
In the signal mode, the fit includes the $\Lb \to \Lz p \bar{p}$ and
$\Xibz \to \Lz p \bar{p}$ signal components together with combinatorial
background.
The signal shapes are modelled with double-sided Crystal Ball functions~\cite{Skwarnicki:1986xj}, while
the combinatorial background is described by exponential functions.
The phase-space-averaged efficiency ratios enter the final fit as
Gaussian-constrained nuisance parameters.

\section{Results}

All results are preliminary. A clear $\Lb \to \Lz p \bar{p}$ signal is observed for the first time.
The statistical significance, computed from the likelihood-ratio test statistic under the null hypothesis of zero signal contribution,
is $5.2\sigma$. After including systematic uncertainties, the significance is
reduced slightly to $5.1\sigma$. The simultaneous fit yields
\begin{equation}
N(\Lb \to \Lz p \bar{p})=39.3\pm9.7,\qquad
N(\Lb \to \Lz K^+K^-)=640\pm31.
\end{equation}
For the $\Xibz \to \Lz p \bar{p}$ decay, only a small excess is observed, with
a significance of $2.3\sigma$. No branching fraction or upper limit is quoted
for this mode, since that would require a dedicated efficiency determination
for $\Xibz \to \Lz p \bar{p}$ decays and knowledge of the relative hadronisation
fraction of \Xibz and \Lb baryons.

The dominant systematic uncertainties arise from tracking efficiency
(4.3\%), fit modelling (2.8\%), and the limited size of the simulated samples
used for the efficiency determination (1.0\%). Smaller contributions arise
from PID efficiency (0.8\%) and truth matching (0.2\%). The total systematic
uncertainty amounts to 5.3\%.

The resulting branching-fraction ratios are
\begin{equation}
R_{\Lb}^{\rm DD}=(4.80\pm1.91)\times10^{-2},\qquad
R_{\Lb}^{\rm LL}=(5.39\pm1.74)\times10^{-2},
\end{equation}
and the combined result is
\begin{equation}
R_{\Lb}=(5.13\pm1.28_{\rm stat}\pm0.27_{\rm syst})\times10^{-2}.
\end{equation}

This result is obtained in the restricted region
$m(h^+h^-)<2.85$ \gev, excluding contributions from intermediate charmonium
resonances decaying to the $p\bar{p}$ and $K^+K^-$ final states.
Since the branching fraction of the normalisation mode is measured without this
requirement, no absolute branching fraction is quoted for
$\Lb \to \Lz p \bar{p}$.

\section{Conclusions}

This analysis constitutes the first observation of the charmless purely
baryonic decay $\Lb \to \Lz p \bar{p}$ and the first observation of a
charmless purely baryonic decay of a baryon.
Larger datasets from LHCb Run~3 will allow more detailed studies of the two-body
invariant-mass spectra, \CP violation, and searches for the related
$\Xibz \to \Lz p \bar{p}$ decay mode.

\section*{References}
\bibliography{moriond}

@article{PDG2024,
     author    = "Navas, S. and others",
    collaboration = "Particle Data Group",
     title     = "{\href{http://pdg.lbl.gov/}{Review of particle physics}}",
     journal   = "Phys. Rev.",
     year = {2024},
     number = {8},
     volume      = "D110",
     pages     = "030001",
     doi = "10.1103/PhysRevD.110.030001"
}

@article{LHCb-PAPER-2017-005,
      author         = "Aaij, R. and others",
      title          = "{Observation of charmless baryonic decays \mbox{\decay{B_{(s)}^0}{\proton\antiproton h^+ h^{\prime -}}}}",
      collaboration  = "LHCb collaboration",
      year           = "2017",
      report         = "{LHCb-PAPER-2017-005 CERN-EP-2017-052}",
      journal        = "Phys. Rev.",
      volume         = "D96",
      pages          = "051103",
      doi            = "10.1103/PhysRevD.96.051103",
      eprint         = "1704.08497",
      archivePrefix  = "arXiv",
      primaryClass   = "hep-ex",
}

@article{LHCb-PAPER-2017-022,
      author         = "Aaij, R. and others",
      title          = "{First observation of the rare purely baryonic decay \mbox{\decay{B^0}{\proton\antiproton}}}",
      collaboration  = "LHCb collaboration",
      year           = "2017",
      journal        = "Phys. Rev. Lett.",
      volume         = "119",
      pages          = "232001",
      doi            = "10.1103/PhysRevLett.119.232001",
      report         = "{LHCb-PAPER-2017-022 CERN-EP-2017-190}",
      eprint         = "1709.01156",
      archivePrefix  = "arXiv",
      primaryClass   = "hep-ex",
}

@article{LHCb-PAPER-2022-032,
      author         = "Aaij, R. and others",
      title          = "{Measurement of the branching fractions $ B^0 \to\proton\antiproton\proton\antiproton$ and  $B_s^0\to\proton\antiproton\proton\antiproton$}",
      collaboration  = "LHCb collaboration",
      report         = "{LHCb-PAPER-2022-032, CERN-EP-2022-206}",
      eprint         = "2211.08847",
      archivePrefix  = "arXiv",
      primaryClass   = "hep-ex",
      year           = "2023",
      journal        = "Phys. Rev. Lett.",
      volume         = "131",
      pages          = "091901",
      doi            = "10.1103/PhysRevLett.131.091901",
}

@article{LHCb-PAPER-2025-032,
      author         = "Aaij, R. and others",
      title          = "{First observation of the charmless baryonic decay $B^+\to\bar{\Lambda}p\bar{p}p$}",
      collaboration  = "LHCb collaboration",
      report         = "{LHCb-PAPER-2025-032, CERN-EP-2025-174}",
      eprint         = "2508.16259",
      archivePrefix  = "arXiv",
      primaryClass   = "hep-ex",
      journal        = "{Phys. Rev. Lett.}",
      volume         = "135",
      pages          = "261901",
      year           = "2025",
      doi            = "10.1103/3pcs-dxtn"
}

@article{LHCb-PAPER-2025-053,
      author         = "Aaij, R. and others",
      title          = "{First observation of the $\bar{B}_s^0\to\Lc\bar{\Lambda}_c^-$ decay and evidence for the $\bar{B}^0\to\Lc\bar{\Lambda}_c^-$ decay}",
      collaboration  = "LHCb collaboration",
      report         = "{LHCb-PAPER-2025-053, CERN-EP-2025-258}",
      eprint         = "2511.20476",
      archivePrefix  = "arXiv",
      primaryClass   = "hep-ex",
      journal        = "{Phys. Rev. Lett.}",
      volume         = "136",
      pages          = "061802",
      year           = "2026",
      doi            = "10.1103/cxn3-8t4g"
}

@article{LHCb-DP-2008-001,
      author         = "Alves~Jr., A. A. and others",
      title          = "{The \lhcb detector at the LHC}",
      collaboration  = "LHCb collaboration",
      journal        = "JINST",
      volume         = "3",
      pages          = "S08005",
      doi            = "10.1088/1748-0221/3/08/S08005",
      year           = "2008",
      report         = "LHCb-DP-2008-001"
}

@article{LHCb-DP-2014-002,
      author         = "Aaij, R. and others",
      title          = "{LHCb detector performance}",
      collaboration  = "LHCb collaboration",
      journal        = "Int. J. Mod. Phys.",
      volume         = "A30",
      pages          = "1530022",
      doi            = "10.1142/S0217751X15300227",
      year           = "2015",
      eprint         = "1412.6352",
      archivePrefix  = "arXiv",
      primaryClass   = "hep-ex",
      report         = "LHCB-DP-2014-002, CERN-PH-EP-2014-290",
}

@article{LHCb-DP-2012-004,
      author         = "Aaij, R. and others",
      title          = "{The \lhcb trigger and its performance in 2011}",
      journal        = "JINST",
      volume         = "8",
      pages          = "P04022",
      doi            = "10.1088/1748-0221/8/04/P04022",
      year           = "2013",
      eprint         = "1211.3055",
      archivePrefix  = "arXiv",
      primaryClass   = "hep-ex",
      report         = "LHCb-DP-2012-004",
}

@article{Sjostrand:2007gs,
      author         = {Sj\"{o}strand, Torbj\"{o}rn and Mrenna, Stephen and
                        Skands, Peter"},
      title          = "{A brief introduction to PYTHIA 8.1}",
      journal        = "Comput. Phys. Commun.",
      volume         = "178",
      pages          = "852-867",
      doi            = "10.1016/j.cpc.2008.01.036",
      year           = "2008",
      eprint         = "0710.3820",
      archivePrefix  = "arXiv",
      primaryClass   = "hep-ph",
      reportNumber   = "CERN-LCGAPP-2007-04, LU-TP-07-28,
                        FERMILAB-PUB-07-512-CD-T",
}

@article{LHCb-PROC-2010-056,
      author         = "Belyaev, I. and others",
      title          = "{Handling of the generation of primary events
                         in Gauss, the LHCb simulation framework}",
      journal="J. Phys. Conf. Ser.",
      volume={331},
      pages={032047},
      doi={10.1088/1742-6596/331/3/032047},
      year={2011},
}

@Article{Lange:2001uf,
     author    = "Lange, D. J.",
     title     = "{The EvtGen particle decay simulation package}",
     journal   = "Nucl. Instrum. Meth.",
     volume    = "A462",
     year      = "2001",
     pages     = "152-155",
     doi       = "10.1016/S0168-9002(01)00089-4",
}

@article{davidson2015photos,
	author="Davidson, N. and Przedzinski, T. and Was, Z.",
	title = "{PHOTOS interface in C++: Technical and physics documentation}",
      	eprint={1011.0937},
      	archivePrefix={arXiv},
      	primaryClass={hep-ph},
	journal = {Comput. Phys. Commun.},
	volume = {199},
	pages = {86},
	year = {2016},
	doi = {https://doi.org/10.1016/j.cpc.2015.09.013},
}

@article{Allison:2006ve,
      author         = "Allison, J. and Amako, K. and Apostolakis, J. and
                        Araujo, H. and Dubois, P.A. and others",
 collaboration = "Geant4 collaboration",
      title          = "{Geant4 developments and applications}",
      journal        = "IEEE Trans.Nucl.Sci.",
      volume         = "53",
      pages          = "270",
      doi            = "10.1109/TNS.2006.869826",
      year           = "2006",
      reportNumber   = "SLAC-PUB-11870",
}

@article{LHCb-DP-2013-002,
      author         = "Aaij, R. and others",
      title          = "{Measurement of the track reconstruction efficiency at LHCb}",
      collaboration  = "LHCb collaboration",
      journal        = "JINST",
      volume         = "10",
      pages          = "P02007",
      doi            = "10.1088/1748-0221/10/02/P02007",
      year           = "2015",
      eprint         = "1408.1251",
      archivePrefix  = "arXiv",
      primaryClass   = "hep-ex",
      report         = "CERN-LHCB-DP-2013-002",
}

@article{LHCb-DP-2018-001,
      author         = "Aaij, R. and others",
      title          = "{Selection and processing of calibration samples to measure the particle identification performance of the LHCb experiment in Run 2}",
      eprint         = "1803.00824",
      archivePrefix  = "arXiv",
      primaryClass   = "hep-ex",
      report         = "LHCb-DP-2018-001",
      year           = "2019",
      journal        = "Eur. Phys. J. Tech. Instr.",
      volume         = "6",
      pages          = "1",
      doi            = "10.1140/epjti/s40485-019-0050-z",
}

@article{Pivk:2004ty,
      author         = "Pivk, Muriel and Le Diberder, Francois R.",
      title          = "{sPlot: A statistical tool to unfold data distributions}",
      journal        = "Nucl. Instrum. Meth.",
      volume         = "A555",
      pages          = "356-369",
      doi            = "10.1016/j.nima.2005.08.106",
      year           = "2005",
      eprint         = "physics/0402083",
      archivePrefix  = "arXiv",
      primaryClass   = "physics.data-an",
      reportNumber   = "LAL-04-07",
}

@article{PBD1,
      author         = "Hsiao, Y. K. and Geng, C. Q. and Rodrigues, E.",
      title          = "{Baryon decays to purely baryonic final states}",
      journal        = "Sci. Rep.",
      volume         = "9",
      pages = "1358",
      doi            = "10.1038/s41598-018-37743-9",
      year           = "2019",
      eprint         = "1806.00861",
      archivePrefix  = "arXiv",
      primaryClass   = "hep-ph",      
}

@article{PhysRevD.72.052002,
  title = {Amplitude analysis of the decay ${B}^{\pm}\to {\pi}^{\pm}{\pi}^{\pm}{\pi}^{\mp}$},
  author={BABAR Collaboration},
  collaboration = {BABAR Collaboration},
  journal = {Phys. Rev.},
  volume = {D72},
  issue = {5},
  pages = {052002},
  numpages = {14},
  year = {2005},
  month = {Sep},
  publisher = {American Physical Society},
  doi = {10.1103/PhysRevD.72.052002},
  url = {https://link.aps.org/doi/10.1103/PhysRevD.72.052002},
  eprint         = "hep-ex/0507025",
  archivePrefix  = "arXiv",
  primaryClass   = "hep-ex",
}

@inproceedings{XGBoosting,
 author = {Chen, Tianqi and Guestrin, Carlos},
 title = {{XGBoost}: A scalable tree boosting system},
 booktitle = {Proceedings of the 22nd ACM SIGKDD International Conference on Knowledge Discovery and Data Mining},
 series = {KDD '16},
 year = {2016},
 isbn = {978-1-4503-4232-2},
 location = {San Francisco, California, USA},
 pages = {785--794},
 doi = {10.1145/2939672.2939785},
 acmid = {2939785},
 publisher = {ACM},
 address = {New York, NY, USA},
 eprint  = "1603.02754",
 archivePrefix = "arXiv",
 primaryClass = "cs.LG"
}

@PhdThesis{Skwarnicki:1986xj,
  author    = "Skwarnicki, Tomasz",
  title     = "{A study of the radiative cascade transitions between the
                  Upsilon-prime and Upsilon resonances}",
  school = 	 {Institute of Nuclear Physics, Krakow},
  year = 	 {1986},
  note = 	 "{\href{http://inspirehep.net/record/230779/}{DESY-F31-86-02}}",
}

@article{LHCb-PAPER-2018-005,
      author         = "Aaij, R. and others",
      title          = "{Observation of the decay \mbox{\decay{\Lb}{\Lc p \antiproton \pim}}}",
      collaboration  = "LHCb collaboration",
      report         = "{LHCb-PAPER-2018-005 CERN-EP-2018-051}",
      eprint         = "1804.09617",
      archivePrefix  = "arXiv",
      primaryClass   = "hep-ex",
      year           = "2018",
      journal        = "Phys. Lett.",
      volume         = "B784",
      pages          = "101",
      doi            = "10.1016/j.physletb.2018.07.033",
}
\end{document}